\documentclass[twoside,preprintnumbers,amsmath,amssymb,pacs,twocolumn,showpacs,nofootinbib]{revtex4}
\usepackage{amsmath,amssymb,url}
\usepackage{graphicx,pslatex,subfigure}

\usepackage{braket,euscript}
\usepackage{fancyvrb}

\renewcommand{\d}{\ensuremath{\mathrm{d}}}

\renewcommand{\d}{\ensuremath{\mathrm{d}}}

\newcommand{\p}{\partial}

\newcommand{\Q}{\ensuremath{\mathcal{Q}}}

\begin{document}

\title{{\bf The Veneziano ghost, glost and gauge copies in QCD }}
\author{D.~Dudal}
\email{david.dudal@kuleuven-kulak.be}
\affiliation{KU Leuven Campus Kortrijk - KULAK, Department of Physics, Etienne Sabbelaan 53, 8500 Kortrijk, Belgium}
\affiliation{Ghent University, Department of Physics and Astronomy, Krijgslaan 281-S9, 9000 Gent, Belgium}
\author{M.~S.~Guimaraes}
\email{msguimaraes@uerj.br}
\affiliation{Instituto de F\'isica Te\'orica, Rua S\~ao Francisco Xavier 524, 20550-013, Maracan\~a, Rio de Janeiro, Brasil}

%\date{\today}
\begin{abstract}
In this short note, we come back to the recent proposal put forward by Kharzeev and Levin \cite{Kharzeev:2015xsa}, in which they phenomenologically couple the non-perturbative Veneziano ghost to the perturbative gluon, leading to a modified gluon propagator (the ``glost'') of the Gribov type, with complex poles. As such, a possible link was made between the QCD topological $\theta$-vacuum (Veneziano ghost) and color confinement (no physically observable gluons). We discuss some subtleties concerning gauge (BRST) invariance of this proposal, related to the choice of Feynman gauge.  We furthermore provide an example in the Landau gauge of a similar phenomenological vertex that also describes the necessary Veneziano ghost but does not affect the Landau gauge gluon propagator.
\end{abstract}
\maketitle
In the recent Letter \cite{Kharzeev:2015xsa}, see also \cite{Kharzeev:2015ifa},  the issue of the Veneziano ghost \cite{Veneziano:1979ec} was revisited. In the absence of massless quarks, the QCD action can explicitly depend on an extra parameter, the $\theta$-angle, closely related to the non-trivial topological nature of the QCD vacuum (cf.~instanton dynamics describing the tunneling between different states with different winding number, \cite{'tHooft:1986nc,Shuryak:1988ck}).

For the sake of presentation, we will mostly follow the notations of \cite{Kharzeev:2015xsa}. A fundamental ingredient in the whole discussion is the topological susceptibility $\chi^4$, defined in the \emph{pure} Yang-Mills case as the zero momentum correlator \cite{Witten:1979vv}
\begin{eqnarray}\label{eq0}
% \nonumber to remove numbering (before each equation)
  i\int \d^4x \braket{\Q(x) \Q(0)} &=& -\chi^4\,,
\end{eqnarray}
where $\Q(x)$ is a pseudoscalar quantity, given by
\begin{equation}\label{eq2}
  \mathcal{Q}(x)=\frac{g^2}{32\pi^2} F_{\mu\nu}(x)\tilde F^{\mu\nu}(x)\,,
\end{equation}
with $\tilde F^{\mu\nu}=\frac{1}{2}\epsilon^{\mu\nu\alpha\beta}F_{\alpha\beta}$ the dual field strength. We refer to \cite{Kharzeev:2015ifa,Kharzeev:2015xsa,Witten:1979vv,Diakonov:1981nv,Hata:1980hn} for more details.  The existence of $\chi^4$ received numerous lattice confirmations, see e.g.~\cite{Campostrini:1988cy,DelDebbio:2004ns,Cichy:2014qta}. The interesting part is that the gauge invariant quantity $\mathcal{Q}(x)$ can be written as a total derivative of a gauge variant pseudovector $\mathcal{K}_\mu(x)$,
\begin{equation}\label{eq3}
  \mathcal{Q}(x)=\p_\mu \mathcal{K}^\mu\,,\mathcal{K}^\mu = \frac{g^2}{16\pi^2}\epsilon^{\mu\nu\rho\sigma}A_\nu^a\left(\p_\rho A_\sigma^a+\frac{g}{3}f^{abc}A_\rho^b A_\sigma^c\right)\,.
\end{equation}
A non-vanishing $\braket{\Q \Q}$ correlator at zero momentum, that is, a non-vanishing topological susceptibility, is thus only possible if there is a massless pole in the $\braket{\mathcal{K}\mathcal{K}}$ correlator given that $\Q$ is a total derivative.  Such a pseudovector pole was introduced first by Veneziano \cite{Veneziano:1979ec}, inspired by Witten \cite{Witten:1979vv}. They were then able to connect the topological susceptibility to the mass of the $\eta'$ particle. More precisely, Veneziano proposed
\begin{equation}\label{eq4}
  \mathcal{K}_{\mu\nu}(q)=i \int \d^4x e^{iqx}\braket{\mathcal{K}_\mu(x) \mathcal{K}_\nu(0)}\stackrel{q^2\sim0}{\sim}  -\frac{\chi^4}{q^2}g_{\mu\nu}\,.
\end{equation}
The negative sign of residue of the massless pole indicates the new ``particle'' is indeed a ghost, see also \cite{Seiler:1987ig}.

Kharzeev and Levin then recognized that the current correlator \eqref{eq4} can be interpreted as being sourced by an effective interaction between the glue and the Veneziano ghost. They postulated such a vertex, and then solved the Dyson-Schwinger equation using solely this coupling, leading to a dynamically corrected gluon propagator (the ``glost'') given by
\begin{equation}\label{eq5}
  G_{\mu\nu}(p^2)=\frac{p^2}{p^4+\chi^4}g_{\mu\nu}\,.
\end{equation}
In the following we will first show that this result necessarily breaks the perturbative BRST symmetry of the underlying theory.

They relied on the Feynman gauge to facilitate computations.  Let us first look again at the Faddeev-Popov action in a general linear covariant gauge, written as\footnote{We switch to Euclidean space time here.}
\begin{equation}\label{4}
    S=\int \d^4x \left(\frac{1}{4}F_{\mu\nu}^2+b^a\p_\mu A_\mu^a +\overline c^a \p_\mu D_\mu^{ab} c^b - \frac{\alpha}{2} b^a b^a\right)
\end{equation}
with covariant derivative
\begin{equation}\label{4d}
    D_\mu^{ab}=\delta^{ab}\p_\mu -gf^{abc}A_\mu^c\,.
\end{equation}
The auxiliary field $b^a$ enforces the linear covariant gauge, with $\alpha=1$ corresponding to the Feynman gauge.

We recall here that the local gauge invariance gets replaced, after gauge fixing, by the BRST invariance \cite{Becchi:1975nq}, a crucial concept to treat gauge theories at the quantum level, used in proofs of perturbative unitarity, renormalizability or quantum gauge invariance.  The BRST variation reads
\begin{equation}\label{4c}
    s A_\mu^a=-D_\mu^{ab} c^b\,,\quad sc^a = \frac{g}{2}f^{abc} c^b c^c\,,\quad  s\overline c^a = b^a\,,\quad sb^a=0\,.
\end{equation}
Hence
\begin{equation}\label{4b}
   S=\int \d^4x \left(\frac{1}{4}F_{\mu\nu}^2\right)+ s \int \d^4x \left (\overline c^a\p_\mu A_\mu^a -\frac{\alpha}{2}b^a\overline c^a\right)\,.
\end{equation}
The tree level gluon propagator in momentum (Fourier) space is easily found, viz.~
\begin{equation}\label{2}
  D_{\mu\nu}(p)=\frac{g_{\mu\nu}}{p^2}-(1-\alpha)\frac{p_\mu p_\nu}{p^4}\,.
\end{equation}
The longitudinal part is thus given by
\begin{equation}\label{3}
   D_L(p)=\frac{\alpha}{p^2}
\end{equation}
and this is in fact an exact result, as a result of BRST symmetry. One way to appreciate this would be to recall that the gluon self-energy corrections are transverse due to BRST invariance, so the longitudinal sector of its inverse (viz.~the gluon propagator) cannot receive any corrections and thus it stays bare. However, as we will discuss later, sometimes one has to be careful when using the gluon self-energy.

Another explicit way to understand \eqref{3}, not making using of the gluon self-energy but still making use of BRST invariance, goes as follows: we can always exactly compute the $b$-propagator from the action \eqref{4}; with $\varphi$ representing any other field present in the action and adding to the action the term $\int \d^4 x J^a b^a $, with $J^a$ an external current;
\begin{equation}\label{6}
\braket{b(x) b(y)}=\left.\frac{\delta^2}{\delta J(y) \delta J(x)}\int [\d\varphi \d b]e^{-S}\right\vert_{J=0}\,.
\end{equation}
We can integrate exactly over the $b$-field, giving
\begin{equation}\label{5}
     \int [\d\varphi db]e^{-S}= \int [\d\varphi]e^{-\int \d^4x\left(\frac{1}{2\alpha} (\p A)^2+\frac{1}{\alpha}J\p A+\frac{J^2}{2\alpha}+\text{rest}\right)}\,.
\end{equation}
Using \eqref{6}, this leads to the exact identification
\begin{equation}\label{7}
   \braket{b(x) b(y)}=\frac{1}{\alpha^2}\braket{\p A(x) \p A(y)}-\frac{\delta(x-y)}{\alpha}\,.
\end{equation}
Since the l.h.s.~must be zero assuming BRST invariance ($\braket{s(\overline c b)}=\braket{bb}$), so must be the r.h.s. As we only relied on the definition of the linear covariant gauge, next to BRST invariance, we can safely state that, according to \eqref{7}, the gluon propagator in that gauge must be of the form
\begin{equation}\label{8}
 D_{\mu\nu}(p)=\Delta(p^2)P_{\mu\nu}(p)+\alpha\frac{p_\mu p_\nu}{p^4}
\end{equation}
in Fourier space, where all non-trivial (non-)perturbative information is collected in the form factor $\Delta(q^2)$ coupled to the transverse projector $P_{\mu\nu}(p)=g_{\mu\nu}-\frac{p_\mu p_\nu}{p^2}$.

Returning to the glost, it is immediately clear, given the incompatibility of the general form \eqref{8} and the glost-result \eqref{eq5}, that the BRST symmetry must be violated, given that the Feynman gauge definition/choice was not explicitly violated.

Next, let us have a closer look on the alluded connection with the Gribov gauge fixing ambiguity. This was first considered by Gribov in the seminal work \cite{Gribov:1977wm} at leading order, later on generalized to all orders by Zwanziger, see \cite{Zwanziger:1989mf} and the recent review \cite{Vandersickel:2012tz}. These works were focused on either the Landau gauge ($\alpha\to0$ limit of the linear covariant gauge, corresponding to $\p A=0$) or the Coulomb gauge, although it was shown later on by Singer that the Gribov problem is of a quite generic nature \cite{Singer:1978dk}. Only more recently, the Gribov issue was considered for the linear covariant gauge and studied explicitly \cite{Sobreiro:2005vn,Capri:2015pja,Capri:2015ixa}. Let us provide a sketchy overview of the Gribov problem. Assuming a linear gauge fixing of the type
\begin{equation}\label{g1}
  \p_\mu A_\mu^a=\alpha b^a\,,
\end{equation}
then the Faddeev-Popov construction implicitly relies on the fact that each gauge orbit only intersects once with the hypersurface defined by \eqref{g1}. Though, Gribov showed this to be wrong at least for $\alpha=0$ (Landau gauge). Assuming an infinitesimally gauge equivalent configuration
\begin{equation}\label{g2}
  \tilde A_\mu^a = A_\mu^a + D_\mu^{ab}\omega^b\,,
\end{equation}
then $\tilde A_\mu^a$ can fulfill the same gauge condition \eqref{g1} iff
\begin{equation}\label{g3}
  -\p_\mu D_\mu^{ab}\omega^b=0\,,
\end{equation}
that is, there are gauge copies if the Faddeev-Popov operator, $-\p D$, has non-trivial zero modes.   In the case of the Landau gauge, this operator is Hermitian, so its eigenvalues are real. Gribov and Zwanziger therefore suggested to restrict the path integration to a subset of gauge configurations, namely those for which the Faddeev-Popov operator is positive. They managed to implement this restriction explicitly into the partition function. This amounts to the introduction of a dynamical mass scale $\lambda^4$ into the theory obeying a self-consistent gap equation, and which affects the gluon propagator. At leading order, one finds \cite{Gribov:1977wm,Zwanziger:1989mf}
\begin{equation}\label{eq10}
  G_{\mu\nu}(p)=\frac{p^2}{p^4+\lambda^4}P_{\mu\nu}(p)\,.
\end{equation}
The presence of the complex conjugate poles in \eqref{eq10} is why Gribov propagators are frequently used in describing confined degrees of freedom, since these cannot have a physical particle (K\"all\'{e}n-Lehmann) interpretation.

The glost propagator \eqref{eq5} thus appears of the Gribov type, \eqref{eq10}, if we identify the Gribov mass scale $\lambda^4$ with the topological susceptibility $\chi^4$. Note that the derivation of the Gribov propagator \eqref{eq10} is \emph{specific} for the Landau gauge, so it does not apply to the Feynman gauge. The generalization of the Gribov-Zwanziger construction to the linear covariant gauges was discussed in \cite{Sobreiro:2005vn,Capri:2015pja,Capri:2015ixa}, leading to a propagator\footnote{The propagators appearing in \cite{Capri:2015pja,Capri:2015ixa} are more general, taking into account further vacuum corrections, which are however unessential for the discussion here.}
\begin{equation}\label{eq11}
   G_{\mu\nu}(p)=\frac{p^2}{p^4+\lambda^4}P_{\mu\nu}(p) + \alpha\frac{p_\mu p_\nu}{p^4}\,.
\end{equation}
As such, the identification of the Feynman gauge glost propagator \eqref{eq5} as the one capable of dynamically curing the Gribov ambiguity based on the topological properties of the QCD vacuum (non-zero $\chi^4$) does not necessarily appear to be well founded. Interestingly, the (partial) resolution of the gauge fixing ambiguity by restricting the integration region of the gauge fields \`{a} la Gribov-Zwanziger\footnote{For an alternative approach, also displaying a soft BRST breaking, see \cite{Serreau:2012cg}.} does lead to a (soft) breaking of the ``standard'' BRST symmetry generated by \eqref{4c}, see for instance \cite{Zwanziger:1989mf,Capri:2015ixa,Dudal:2008sp,Schaden:2014bea}. Though, a non-perturbative BRST symmetry can be introduced, depending on the Gribov mass $\lambda^4$, which allows, for example, to still prove the non-renormalization of the longitudinal gluon propagator. Next to analytical arguments \cite{Capri:2015ixa,Huber:2015ria,Aguilar:2015nqa}, this has also been explicitly observed in lattice simulations \cite{Cucchieri:2009kk,Bicudo:2015rma}, which in principle encompass the topological raison d'\^{e}tre of $\braket{\Q\Q}$. It is important to mention that, when using the Gribov-Zwanziger formalism, additional fields are required to maintain locality \cite{Zwanziger:1989mf,Vandersickel:2012tz}. These extra fields enter the BRST transformation and mix up with e.g.~the gauge field. As such, it makes no longer sense to work with only the gluon self energy, since it will be part of a 1PI matrix that defines, after inversion, the (connected) propagator matrix. For an explicit example, see \cite{Gracey:2009mj}.

As the Gribov issue has been studied to more extent in the Landau gauge, we should not refrain from carrying out a similar exercise as \cite{Kharzeev:2015xsa} in this particular gauge. We first notice that \eqref{eq4} is not the unique sensible way to introduce the Veneziano ghost. The most important ingredient is to have the zero mass pole, so
\begin{equation}\label{eq50}
  \mathcal{K}_{\mu\nu}(q)=i \int \d^4x e^{iqx}\braket{\mathcal{K}_\mu(x) \mathcal{K}_\nu(0)}\stackrel{q^2\sim0}{\sim} -\frac{\chi^4}{q^2}\frac{q_\mu q_\nu}{q^2}
\end{equation}
will equally well do the job (see also \cite{Diakonov:1981nv,Zhitnitsky:2013hs}), i.e.~without changing the gauge invariant $\braket{\Q\Q}$-correlator \eqref{eq0}. Any suitably chosen linear combination of \eqref{eq4} and \eqref{eq50} is also possible. Adding a term transversal in $q$ will  not affect $\braket{\Q\Q}$ either. We will now benefit from these observations. Following \cite{Kharzeev:2015xsa}, from \eqref{eq50} we may define an effective infrared Veneziano ghost-gluon-gluon vertex $\Gamma_\mu^{\alpha\beta}(q,p)$,
\begin{eqnarray}\label{eq51}
  \frac{1}{(2\pi)^4i} \int \d^4p  \Gamma_\mu^{\alpha\beta} \Gamma_\nu^{\rho\sigma} \frac{P_{\alpha\rho}(p)}{p^2}\frac{P_{\beta \sigma}(p-q)}{(p-q)^2}=-\frac{\chi^2}{q^2}\frac{q_\mu q_\nu}{q^2}\,,
\end{eqnarray}
where we assumed the Landau gauge condition. From the Landau gauge defining Ward identity $\frac{\delta \Gamma}{\delta b}= \partial_\mu A_\mu$, with $\Gamma$ the 1PI generating functional, it can be easily shown that the Landau gauge gluon propagator is necessarily transverse, thus proportional to the already defined projector $P_{\mu\nu}$.

A (possible) solution to \eqref{eq51} is provided by
\begin{eqnarray}\label{eq52}
  &&\Gamma_\mu^{\alpha\beta}(q,p) \propto X q_\mu (p-q)^{\alpha}q^\beta\;;\quad q\leq p \nonumber\\X^2&=&\frac{-\chi^2}{p^2 q^2}\frac{1}{(p-q)_\alpha (p-q)_\rho q_\beta q_\sigma P^{\alpha\rho}(q)P^{\beta\sigma}(p-q)}\,,
\end{eqnarray}
where we omitted some numerical prefactors irrelevant for further purposes. The indices $\mu$, resp.~$\alpha,\beta$ refer to the Veneziano ghost, resp.~gluons.

Assuming the infrared physics is dominated by the glost\footnote{That means, we will omit all information from the already present quark, gluon and ghost vertices, see also \cite{Kharzeev:2015xsa,Kharzeev:2015ifa}.}, we can compute the one loop gluon self energy $\Sigma$ using the vertex $\Gamma_\mu^{\alpha\beta}$. The diagram shown in FIG.~1 leads to
\begin{eqnarray}\label{int}
% \nonumber to remove numbering (before each equation)
  \Sigma_{\mu\nu}(p) &=& \frac{1}{(2\pi)^4i}\int \d^4q \Gamma_\alpha^{\mu\beta}(q,p) \frac{q_{\alpha}q_{\rho}}{q^4}\frac{P_{\beta\sigma}(p-q)}{(p-q)^2}\Gamma_{\rho}^{\nu\sigma}(q,p)\,,\nonumber\\
\end{eqnarray}
where we used for the internal Veneziano ghost a propagator of the form $\frac{q_\alpha q_{\rho}}{q^4}$, consistent with our earlier choice \eqref{eq51}.
\begin{figure}[t]
\label{selfenergy}
       \centering  % figura centralizada
       \includegraphics[scale=0.5]{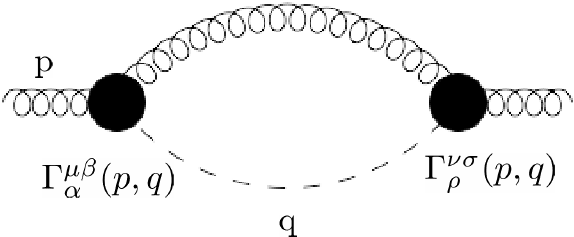}
       \caption{One loop gluon self energy: dashed line stands for the Veneziano ghost propagator, curly line represents the perturbative gluon, the black dots are vertices.}
\end{figure}
As in \cite{Kharzeev:2015xsa}, we can approximate the integral \eqref{int} by assuming $q\ll p$ as we are considering non-perturbative effects in the very deep infrared. Because of our choice of vertex \eqref{eq52}, we would find, for $q\ll p$,
\begin{equation}
  \Gamma_\alpha^{\mu\beta}(q,p)\Gamma_{\rho}^{\nu\sigma}(q,p)\stackrel{q\ll p}\propto p_{\mu} p_\nu
\end{equation}
and a fortiori, we would thus have
\begin{eqnarray}\label{intbis}
% \nonumber to remove numbering (before each equation)
  \Sigma_{\mu\nu}(p) &\propto& p_{\mu} p_\nu\,,
\end{eqnarray}
i.e.~the gluon self energy becomes longitudinal in this case. It is then immediate that the Landau gauge glost propagator will coincide with the perturbative gluon one, since the transverse projector, that is by definition present in the propagator, will always project to zero the self-energy correction \eqref{intbis}.

Notice that we are not proclaiming that the vertex defined via \eqref{eq51}-\eqref{eq52} is the correct one in the Landau gauge\footnote{Rather the contrary, since the eventual gluon self energy is not transverse either.}, as other possibilities exist, as long as the gauge invariant correlation function $\braket{\Q\Q}$ is correctly reproduced. We merely wanted to illustrate that depending on the vertex choice, quite some different glost propagator dynamics could emerge. However, to find out which vertex is effectively realized per gauge is open for future debate. Another strong constraint, next to consistency with $\braket{\Q\Q}$, is that the longitudinal part of the gluon propagator in the class of linear covariant gauges should receive no quantum corrections. In brief, a possible strategy to proceed would be to set
\begin{equation}\label{eq50tris}
  \mathcal{K}_{\mu\nu}(q)\stackrel{q^2\sim0}{\sim} -\frac{\chi_1^4}{q^2}\delta_{\mu\nu}-\frac{\chi_2^4}{q^2}\frac{q_\mu q_\nu}{q^2}+f(p,q)P_{\mu\nu}(q)\,,
\end{equation}
with $\chi^4=\chi_1^4+\chi_2^4$; and then to propose Ans\"atze for the vertex $\Gamma_\mu^{\alpha\beta}(q,p)$ such that \eqref{eq50tris} can be realized in conjunction with a longitudinal gluon propagator projection that remains $\frac{\alpha}{p^2}$.

At last, it is worth mentioning that the Gribov problem has been extensively studied in the Coulomb gauge as well, in which case \cite{Gribov:1977wm,Zwanziger:2004np} a gluon propagator
\begin{equation}\label{coul}
  D(\vec{p},p_0)=\frac{\vec{p}^2}{(p_0^2+\vec{p}^2)\vec{p}^2+M^4}
\end{equation}
was predicted (up to a projector), a form that is indeed pretty consistent with corresponding lattice data \cite{Burgio:2008jr}. It would appear that \eqref{coul} does not entirely match the Coulomb gauge glost propagator used in \cite{Kharzeev:2015xsa} to analyze a non-perturbative infrared coupling constant. In \cite{Burgio:2009xp}, a link was also made with BRST breaking.

In summary, we have provided some arguments why it appears to be premature to directly link the topological nature of the QCD vacuum to confinement, or more precisely, to link the Veneziano ghost to the issue of Gribov copies and confinement. We discussed the apparent lack of BRST invariance, related to a strong constraint on the longitudinal sector of the gluon/glost propagator that is not fulfilled. As respecting BRST invariance is crucial to extract gauge invariant (or better said, gauge parameter independent) physical information, one does not have the liberty to tamper too much with effective vertices and/or propagators. Therefore, we paid a somewhat closer view at the connection with the Gribov gauge fixing ambiguity, reporting some different behaviour here than from the introduced gluon/glost propagator. Furthermore, we provided an example of a ``Veneziano vertex'' that does not affect the gluon propagator at all, at least in the Landau gauge.

Though, further research is definitely needed to provide a firmer answer to the premise of \cite{Kharzeev:2015xsa}, and either to confirm or to falsify it. In our opinion, a quite delicate point is the gauge dependent nature of the $\mathcal{K}_\mu$-current and its correlation function. Since this is a genuinely non-perturbative correlation function, it would already be most interesting to have specific information about it using e.g.~\emph{gauge fixed} lattice simulations or perhaps even via functional methods. That such might become possible in the future is not unrealistic, given the recent progress in accessing the linear covariant gauges non-perturbatively \cite{Capri:2015ixa,Huber:2015ria,Aguilar:2015nqa,Cucchieri:2009kk,Bicudo:2015rma,Siringo:2014lva}. This might also put further restrictions on how the effective Veneziano ghost-gluon interaction must be modeled. In any case, lattice Landau gauge is well-matured by now, so that might be the first option to explore the Veneziano ghost in a fully non-perturbative gauge fixed setting.

\section*{Acknowledgments}
We thank A.~Duarte, D.~Kharzeev, E.~Levin, O.~Oliveira, L.~F.~Palhares, P.~J.~Silva and S.~P.~Sorella for discussions. M.~S.~Guimaraes thanks ``The Conselho Nacional de Desenvolvimento Cient\'{i}fico e Tecnol\'{o}gico'' (CNPq-Brazil) for the financial support and gratefully acknowledges the hospitality at the KU Leuven Campus Kortrijk - KULAK where part of this work was executed.

\end{document}